\documentclass[aps,prl,twocolumn,superscriptaddress,nofootinbib]{revtex4-2}
\usepackage{amsmath,amssymb,bm,mathtools}
\usepackage{graphicx}
\usepackage{booktabs}
\usepackage{hyperref}
\hypersetup{hidelinks}
\newcommand{\F}{\mathbb F}

\newcommand{\Rcal}{\mathcal R}
\newcommand{\ketM}{\lvert M_U\rangle}
\newcommand{\ketMa}{\lvert M_{U,a}\rangle}
\newcommand{\wt}{\operatorname{wt}}

\begin{document}

\title{Coded Clifford Measurements for Multiqubit Magic-State Cultivation}

\author{Gunsik Min}
\email{mgs3351@korea.ac.kr}
\affiliation{School of Electrical Engineering, Korea University, Seoul, Republic of Korea}

\author{Jun Heo}
\email{junheo@korea.ac.kr}
\affiliation{School of Electrical Engineering, Korea University, Seoul, Republic of Korea}

\begin{abstract}
Magic-state cultivation suppresses errors by measuring logical Clifford checks and discarding inconsistent outcomes. For an entangled resource, several measurement branches are usable, so their outcomes form a multibit classical record whose corruption can produce a logical-frame error. We show that this measurement record can be protected as a binary linear code. For any third-level Clifford-hierarchy unitary $U\in\mathcal C_3$, every parity of the branch bits can be measured by a commuting Hermitian Clifford check $C(v)=UX(v)U^\dagger$. Choosing which parities to measure therefore defines a binary code $y(a)=aG$. If the valid records have minimum distance $d$, at least $d$ readout-bit flips are required to confuse one valid branch with another, giving $P_{\rm wv}=O(q^d)$. A Plotkin bound limits the length of any binary branch record, and the Clifford schedules attain this limit: at distance four, six logical measurements suffice for $|CS\rangle$ and seven for $|CCZ\rangle$, compared with eight and twelve under independent repetition. Thus restricting the logical schedule to Clifford parities requires no additional measurements. In a native-CZZ-assisted Steane implementation, the $[6,2,4]$ CS schedule is also the unique minimum-cost distance-four solution within the compiled Clifford-check family, reducing the cultivation core by $26.6\%$ in active locations. State-vector simulations without a final ideal code-space projection further show higher acceptance and approximately half the residual error weight on the two Steane blocks. Coding the logical measurement record therefore reduces both measurement redundancy and compiled fault-tolerant overhead.
\end{abstract}

\maketitle

High-fidelity non-Clifford resource states remain a major overhead in fault-tolerant quantum computation, motivating extensive work on magic-state distillation, factory design, and low-overhead resource preparation~\cite{BravyiKitaev2005,BravyiHaah2012,Meier2013,Jones2013Multilevel,Haah2017Magic,HastingsHaah2018,Litinski2019Magic,OGormanCampbell2017,GidneyFowler2019,CampbellTerhalVuillot2017}. Surface-code and lattice-surgery architectures make this resource cost especially explicit~\cite{Dennis2002,Fowler2012,Horsman2012,Litinski2019Game}. Magic-state cultivation has emerged as a low-overhead alternative to multilevel distillation, with surface-code, color-code, and constant-depth variants under active development~\cite{Gidney2024,Chen2026,Vaknin2026,Sahay2026,Hetenyi2026}; cultivation has also been demonstrated on a superconducting processor together with code switching~\cite{Rosenfeld2025}. These protocols repeatedly measure logical Clifford symmetries and keep only consistent outcomes. For a one-qubit $T$ resource, the retained branch is labeled by one bit, so repetition is natural. Entangled resources such as $|CS\rangle$ and $|CCZ\rangle$ are also active factory, distillation, and gate-synthesis targets~\cite{Chamberland2022,Gong2026,Eastin2013Toffoli,Jones2013Toffoli,Jones2013Composite,CampbellHoward2017PRL,CampbellHoward2017PRA,HaahHastings2018,GidneyFowler2019}. When several Clifford-related branches are accepted, however, the branch label contains several bits. Corrupting that label produces a logical-frame error on an otherwise accepted magic state, while independently repeating each of the $k$ branch generators $d$ times costs $kd$ logical measurements.

Classical redundancy can do better than independent repetition. Fault-tolerant syndrome extraction and redundant measurements have a long history~\cite{Shor1996,Steane1996,Steane1999,SteaneIbinson2005,Preskill1998,Knill2005,Fujiwara2014,Ashikhmin2020,ChamberlandBeverland2018,ChaoReichardt2018}, and Ouyang showed that arbitrary projective measurements can be protected by assigning classical codewords to their projectors and measuring corresponding commuting spectral observables~\cite{Ouyang2024}. For cultivation, the difficulty is physical: an observable constructed from a classical code need not correspond to a logical check that can be measured fault tolerantly in the chosen quantum code. This is closely related to the broader constraints on transversal and locality-preserving logical gates~\cite{EastinKnill2009,BravyiKoenig2013,PaetznickReichardt2013,Anderson2014,Bombin2015,Kubica2015,Cui2017}. We show that third-level resource bases have a useful closure property: every parity of the branch bits corresponds to a commuting Hermitian Clifford check $C(v)=UX(v)U^\dagger$. These checks can therefore be organized directly as a classical code for the measurement record.

Restricting the schedule to Clifford parities need not require extra measurements. We prove an exact correspondence between parity schedules and binary linear record codes, show that the code distance determines how many readout-bit errors are required to confuse two valid branches, and apply a binary Plotkin bound to all $2^k$ branches. For distance four, the Clifford schedules for $T$, $CS$, and $CCZ$ reach this bound even though a general robust-measurement scheme could assign arbitrary binary records that are not generated by linear parities. The logical measurement count drops from $8$ to $6$ for $|CS\rangle$ and from $12$ to $7$ for $|CCZ\rangle$. For a concrete Steane/native-CZZ realization of the $CS$ checks, we also solve the minimum-cost measurement problem at fixed record distance and find the coded schedule to be the unique minimum-cost solution. Exact state-vector Monte Carlo without a final ideal code-space projection further shows that the shorter schedule raises acceptance and lowers the residual error weight at the output of the Steane stage under the same gate/readout noise model.

Let $\mathcal C_1$ denote the Pauli group and $\mathcal C_3$ the third level of the Clifford hierarchy, whose role in fault-tolerant gate constructions is well established~\cite{GottesmanChuang1999,EastinKnill2009,BravyiKoenig2013,Cui2017}. For $k$ qubits define
\begin{equation}
 \ketM=U\lvert+\rangle^{\otimes k},\qquad U\in\mathcal C_3,
 \label{eq:resource}
\end{equation}
and for $v=(v_1,\ldots,v_k)\in\F_2^k$,
\begin{equation}
 X(v)=\prod_{j=1}^k X_j^{v_j},\qquad C(v)=UX(v)U^\dagger.
 \label{eq:Cv}
\end{equation}
By definition of the hierarchy, $C(v)$ is Clifford. Unitary conjugation of $X(v)$ also gives $C(v)^\dagger=C(v)$ and $C(v)^2=I$. Since the $X(v)$ commute,
\begin{equation}
 C(v)C(w)=C(v+w),\qquad [C(v),C(w)]=0.
 \label{eq:abelian}
\end{equation}
Thus the operators that distinguish the branches form a commuting family of Hermitian Clifford checks with eigenvalues $\pm1$.

Define the orthonormal resource basis
\begin{equation}
 \ketMa=UZ(a)\lvert+\rangle^{\otimes k},\qquad a\in\F_2^k,
 \label{eq:branches}
\end{equation}
with projectors $P_a=\lvert M_{U,a}\rangle\!\langle M_{U,a}\rvert$. The key algebraic fact is
\begin{equation}
 C(v)\ketMa=(-1)^{a\cdot v}\ketMa,
 \label{eq:eigen}
\end{equation}
which follows from $X(v)Z(a)=(-1)^{a\cdot v}Z(a)X(v)$. Equivalently,
\begin{equation}
 C(v)=\sum_{a\in\F_2^k}(-1)^{a\cdot v}P_a.
 \label{eq:spectral}
\end{equation}
The sign $(-1)^{a\cdot v}$ gives the parity of the branch bits selected by $v$. Equation~(\ref{eq:spectral}) is the spectral observable used in the code-based measurement construction of Ref.~\cite{Ouyang2024}; for the present resource basis, this observable is automatically Clifford. This closure turns a classical linear record directly into a cultivation schedule. In addition,
\begin{equation}
 \ketMa=F_a\ketM,\qquad F_a=UZ(a)U^\dagger\in\mathcal C_2.
 \label{eq:frame}
\end{equation}
Hence all ideal branches are Clifford-equivalent. For diagonal $U$, including $T$, $CS$, and $CCZ$, $F_a=Z(a)$, so branch handling reduces to a Pauli-frame update.

\begin{figure*}[t]
\includegraphics[width=\textwidth]{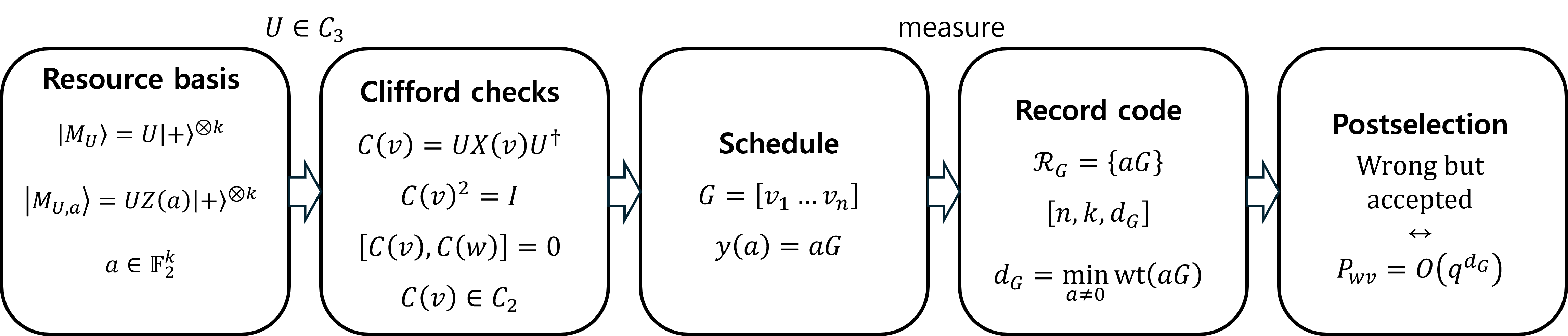}
\caption{Logical-level connection between cultivation and classical coding. For $U\in\mathcal C_3$, each parity of the branch bits is measured by a commuting Clifford check $C(v)$. Choosing check labels $v_t$ defines the measurement schedule $G=[v_1\cdots v_n]$ and the set of valid branch records $\mathcal R_G=\{aG\}$. Postselection rejects records outside this set. If the valid records have minimum distance $d_G$, fewer than $d_G$ flipped record bits cannot change one valid branch record into another, giving $P_{\rm wv}=O(q^{d_G})$.}
\label{fig:correspondence}
\end{figure*}

Measure $n$ Clifford checks $C(v_t)$ and collect their labels as columns of
\begin{equation}
 G=[v_1\ v_2\ \cdots\ v_n]\in\F_2^{k\times n},\qquad \operatorname{rank}G=k.
 \label{eq:G}
\end{equation}
Writing the measured eigenvalue as $(-1)^{y_t}$, Eq.~(\ref{eq:eigen}) gives the ideal record
\begin{equation}
 y(a)=aG.
 \label{eq:record}
\end{equation}
Therefore the $2^k$ ideal records form the binary linear code
\begin{equation}
 \Rcal_G=\{aG:a\in\F_2^k\},\qquad
 d_G=\min_{a\ne0}\wt(aG).
 \label{eq:recordcode}
\end{equation}
Conversely, every full-rank binary matrix $G$ specifies which logical Clifford parities should be measured through Eq.~(\ref{eq:Cv}). This gives an exact logical-level correspondence between the measurement schedule and a classical code for the branch record; the physical cost of measuring a given $C(v)$ remains code- and hardware-dependent.

Cultivation uses postselection rather than classical error correction. Suppose the ideal record is $aG$ while the observed record is $aG+e$, and accept only if the observed record belongs to $\Rcal_G$. By linearity,
\begin{equation}
 aG+e\in\Rcal_G \iff e\in\Rcal_G.
\end{equation}
Thus the observed record corresponds to a different valid branch if and only if
\begin{equation}
  e\in\Rcal_G\setminus\{0\}.
 \label{eq:wrongvalid}
\end{equation}
Every nonzero record error of weight below $d_G$ is therefore rejected. If record bits flip independently with probability $q$ and $A_w$ is the weight enumerator of $\Rcal_G$,
\begin{equation}
 P_{\rm wv}(q)=\sum_{w=d_G}^{n}A_w q^w(1-q)^{n-w}
 =A_{d_G}q^{d_G}+O(q^{d_G+1}).
 \label{eq:Pwv}
\end{equation}
The same exponent follows under local-stochastic record noise. Therefore, for errors confined to the classical measurement record, at least $d_G$ flipped bits are required to accept the wrong branch. This statement is deliberately separate from faults that alter the quantum data during a logical-check gadget. If logical failures caused by data and decoder faults first appear at order $p^\delta$ and record noise satisfies $q=\Theta(p)$, choosing $d_G\ge\delta+1$ makes measurement-record corruption higher order in $p$ without changing the leading data-fault exponent.

The correspondence immediately imports classical bounds, but the most useful lower bound for cultivation has a direct branch-counting proof and does not assume linearity.

\textit{Theorem 1 (minimum binary record length).---}
Any length-$n$ binary record that distinguishes all $M=2^k$ branches with pairwise Hamming distance at least $d$ obeys
\begin{equation}
 n\ge
 \left\lceil d\frac{2^k-1}{2^{k-1}}\right\rceil.
 \label{eq:globalbound}
\end{equation}
At a single coordinate, if $s$ of the $M$ records contain $1$, that coordinate contributes $s(M-s)\le M^2/4$ to the sum of all pairwise distances. Summing over $n$ coordinates and comparing with $d\binom{M}{2}$ proves Eq.~(\ref{eq:globalbound}). This is the Plotkin average-distance bound specialized to exactly $2^k$ binary branch labels~\cite{Plotkin1960}.

When $d=r2^{k-1}$, measuring every nonzero parity $C(v)$, $v\in\F_2^k\setminus\{0\}$, exactly $r$ times saturates the bound and gives the repeated simplex code
\begin{equation}
 [n,k,d]=[r(2^k-1),\ k,\ r2^{k-1}],
 \label{eq:simplex}
\end{equation}
and is therefore globally optimal even if arbitrary nonlinear binary records are allowed. A generic robust-projective-measurement construction may use such nonlinear labels, whereas Eq.~(\ref{eq:Cv}) guarantees a Clifford observable for every linear parity. When Eq.~(\ref{eq:simplex}) reaches the same lower bound, restricting the schedule to Clifford parities requires no additional measurements.

At target distance four this gives
\begin{center}
\begin{tabular}{c c c c}
\toprule
resource & independent repetition & coded schedule & reduction \\
\midrule
$|T\rangle$   & $4$  & $[4,1,4]$ & $0\%$ \\
$|CS\rangle$  & $8$  & $[6,2,4]$ & $25\%$ \\
$|CCZ\rangle$ & $12$ & $[7,3,4]$ & $42\%$ \\
\bottomrule
\end{tabular}
\end{center}
For $k=1$ the optimum is ordinary repetition. For $k>1$, products of commuting Clifford symmetries carry parity information about several branch bits at once and compress the redundancy.

The $k=2$ case shows explicitly how parity coding removes redundant logical measurements. A complete fault-tolerant realization of the six-check $|CS\rangle$
schedule, including verified CAT7 readout, Steane error detection,
surface-code growth, and end-to-end decoding, is developed separately
in Ref.~\cite{MinHeo2026DirectCS}; here $|CS\rangle$ serves as a concrete
example of the general record-code correspondence. For $U=CS_{AB}$,
\begin{align}
 C_1&=X_A S_B CZ_{AB},\qquad C_2=X_B S_A CZ_{AB},\nonumber\\
 C_{12}&=C_1C_2,
 \label{eq:CSchecks}
\end{align}
and the six-check schedule
\begin{equation}
 G_{CS}=\begin{pmatrix}
 1&0&1&1&0&1\\
 0&1&1&0&1&1
 \end{pmatrix}
 \label{eq:G6}
\end{equation}
is a $[6,2,4]$ record. The best five-check schedule has distance three. Adding the sixth parity check raises the minimum number of readout-bit flips required to confuse two valid branches from three to four, changing $P_{\rm wv}$ from $O(q^3)$ to $O(q^4)$; the exact weight enumerators are given in the Supplemental Material~\cite{SM}. More importantly for physical cost, a native-CZZ-assisted Steane compilation assigns $98$ active locations to each $C_1$ or $C_2$ measurement and $70$ to $C_{12}$, including its verified CAT7 ancilla. Let $(m_1,m_2,m_{12})$ denote their multiplicities. Distance four is equivalent to
\begin{equation}
 m_1+m_{12}\ge4,\quad m_2+m_{12}\ge4,\quad m_1+m_2\ge4.
 \label{eq:weightedCS}
\end{equation}
Minimizing $98(m_1+m_2)+70m_{12}$ over nonnegative integers gives the unique optimum $(2,2,2)$, with measurement-stage cost $532$. Adding the fixed $164$-location two-block midpoint detector gives $696$ active locations, versus $948$ for independent quartic repetition: a $26.6\%$ reduction. An independent qubit-step estimate gives $2040$ versus $2844$ active-qubit gate-steps, a $28.3\%$ reduction~\cite{SM}. Thus the six-check record is both shortest and minimum cost among the compiled Clifford-parity schedules considered here; neither metric is a full factory spacetime cost.

For $U=CCZ_{123}$,
\begin{equation}
 C_1=X_1CZ_{23},\quad C_2=X_2CZ_{13},\quad C_3=X_3CZ_{12}.
 \label{eq:CCZchecks}
\end{equation}
Measuring the seven nonzero products $C(v)$ once gives
\begin{equation}
 G_{CCZ}=\begin{pmatrix}
1&0&0&1&1&0&1\\
0&1&0&1&0&1&1\\
0&0&1&0&1&1&1
\end{pmatrix},
 \label{eq:GCCZ}
\end{equation}
a $[7,3,4]$ simplex record. Equation~(\ref{eq:globalbound}) gives $n\ge7$, proving that no shorter binary record---linear or nonlinear---can resolve all eight branches at distance four. Thus the Clifford parity family guaranteed by Eq.~(\ref{eq:Cv}) is as short as any binary record, linear or nonlinear.

\begin{figure*}[t]
\centering
\includegraphics[width=0.28\textwidth]{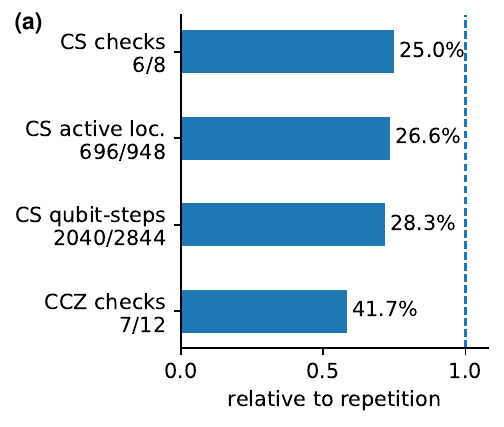}\hfill
\includegraphics[width=0.33\textwidth]{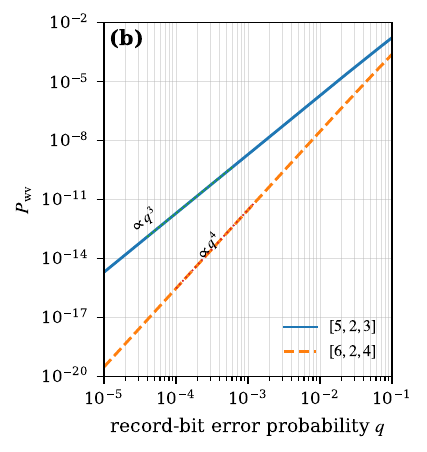}\hfill
\includegraphics[width=0.30\textwidth]{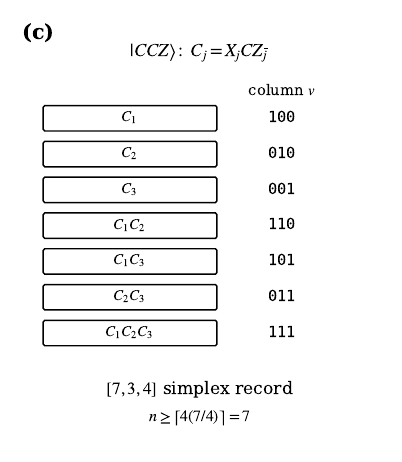}
\caption{Consequences of coded Clifford records. (a) For distance four, coding reduces the $CS$ schedule from $8$ to $6$ logical checks, from $948$ to $696$ compiled-core active locations, and from $2844$ to $2040$ active-qubit gate steps; the $CCZ$ schedule falls from $12$ to $7$ logical checks. (b) The best $[5,2,3]$ and $[6,2,4]$ two-bit records require three and four readout-bit flips, respectively, to confuse two valid branches. (c) The $[7,3,4]$ $CCZ$ simplex schedule reaches the minimum length allowed for any binary record.}
\label{fig:applications}
\end{figure*}

The correspondence is intentionally logical-level. A logical Clifford operator is not automatically cheap to measure fault tolerantly. This distinction between algebraically available logical gates and physically fault-tolerant implementations is central to transversal-gate, gauge-fixing, and code-switching approaches~\cite{EastinKnill2009,PaetznickReichardt2013,Anderson2014,Bombin2015,Kubica2015}. Gauging methods demonstrate fault-tolerant measurement of important logical symmetries in specific codes~\cite{WilliamsonYoder2026,Hetenyi2026}; architecture-specific compilation of the multiblock $CCZ$ checks remains a separate problem. The Supplemental Material gives the exact action of the corresponding projective measurements and makes this scope explicit~\cite{SM}.

The main result is not a new classical code bound, nor the first use of coded quantum measurements. The key result is that the measurements required by a classical record code can be chosen as logical Clifford checks for third-level magic-state branches. In a generic projective measurement, the code-inspired observables of Ref.~\cite{Ouyang2024} can be arbitrary Hermitian operators. For the branch basis $UZ(a)|+\rangle^{\otimes k}$, every parity of the branch bits is instead represented by the logical Clifford symmetry $UX(v)U^\dagger$, and the corresponding branches are Clifford-frame equivalent. The code distance can therefore protect a record of fault-tolerant Clifford measurements directly, rather than a set of generic spectral observables.

The distance-four $CS$ and $CCZ$ examples make this distinction concrete. Even though a general robust-measurement scheme may use nonlinear binary labels, the $[6,2,4]$ and $[7,3,4]$ Clifford schedules already reach the same lower bound. Thus restricting the schedule to Clifford parities requires no additional measurements for these targets. In the compiled $CS$ example, Eq.~(\ref{eq:weightedCS}) makes $(2,2,2)$ the unique hardware-weighted distance-four optimum: $948\to696$ active locations and $2844\to2040$ active-qubit gate steps. A stricter state-vector test removes the final ideal code-space projection and examines the output of the Steane stage directly. At $p=10^{-3}$ the coded schedule accepts $0.7350(62)$ versus $0.6256(68)$, with mean residual Pauli weight $0.0639(54)$ versus $0.1263(84)$ on the two Steane blocks; the full sweep is in the Supplemental Material~\cite{SM}. These quantities characterize the state entering the downstream stage; they are not full-factory logical-error estimates.

Record distance does not by itself set the complete factory distance: gadget faults can change both data and reported outcomes. This suggests a co-design principle: the record code determines which branch parities should be measured and how often, while the QEC code and hardware determine the cost of those measurements. Unequal costs lead to the weighted optimization problem derived in the Supplemental Material~\cite{SM}. Because the correspondence holds for any $U\in\mathcal C_3$, it provides a common design language for $T$, $CS$, $CCZ$, and more general third-level resource states.

\begin{acknowledgments}
\end{acknowledgments}

\end{document}


\title{Supplemental Material for\\``Coded Clifford Measurements for Multiqubit Magic-State Cultivation''}
\author{Gunsik Min}
\affiliation{School of Electrical Engineering, Korea University, Seoul, Republic of Korea}
\maketitle

\section{Algebraic relation to prior coded measurements}
The Letter uses the resource branches
\begin{equation}
 |M_{U,a}\rangle=UZ(a)|+\rangle^{\otimes k},\qquad U\in\mathcal C_3,
\end{equation}
and the parity observables
\begin{equation}
 C(v)=UX(v)U^\dagger,\qquad a,v\in\mathbb F_2^k.
\end{equation}
Because $X(v)$ is Pauli and $U\in\mathcal C_3$, $C(v)$ is Clifford. Unitary conjugation also gives
\begin{equation}
 C(v)^\dagger=C(v),\qquad C(v)^2=I,\qquad [C(v),C(w)]=0.
\end{equation}
Using $X(v)Z(a)=(-1)^{a\cdot v}Z(a)X(v)$,
\begin{equation}
 C(v)|M_{U,a}\rangle=(-1)^{a\cdot v}|M_{U,a}\rangle.
 \label{eq:eigenS}
\end{equation}
Thus $C(v)$ measures the binary branch parity $a\cdot v$. With $P_a=|M_{U,a}\rangle\langle M_{U,a}|$,
\begin{equation}
 C(v)=\sum_{a\in\mathbb F_2^k}(-1)^{a\cdot v}P_a.
 \label{eq:spectralS}
\end{equation}
The ideal branches are Clifford-frame equivalent because
\begin{equation}
 |M_{U,a}\rangle=UZ(a)U^\dagger|M_U\rangle,
 \qquad UZ(a)U^\dagger\in\mathcal C_2.
\end{equation}
For diagonal $U$, including $T$, $CS$, and $CCZ$, the frame is simply $Z(a)$.

Ouyang's robust-projective-measurement construction~\cite{Ouyang2024} assigns classical codewords to projectors and measures commuting spectral observables that reveal the corresponding code symbols. For a binary linear record $aG$, a column $v_t$ of $G$ selects the symbol $a\cdot v_t$. Equation~(\ref{eq:spectralS}) shows that the associated spectral observable is exactly $C(v_t)=UX(v_t)U^\dagger$. The new structural point used in the Letter is therefore the Clifford closure of these code-inspired observables for third-level resource bases, not the abstract idea of coding measurement outcomes.

\section{Additional properties of the measurement record}
Let $G\in\mathbb F_2^{k\times n}$ have rank $k$ and let
\begin{equation}
 \mathcal R_G=\{aG:a\in\mathbb F_2^k\}
\end{equation}
be the valid-record set. If the ideal record is $aG$ and the observed record is $aG+e$, validity postselection gives
\begin{equation}
 \text{wrong but accepted}
 \Longleftrightarrow
 e\in\mathcal R_G\setminus\{0\}.
 \label{eq:wrongS}
\end{equation}
Hence a code of distance $d$ rejects every nonzero record error of weight below $d$. For independent bit flips with probability $q$ and weight enumerator $A_w$,
\begin{equation}
 P_{\rm wv}(q)=\sum_{w=d}^{n}A_wq^w(1-q)^{n-w}
 =A_dq^d+O(q^{d+1}).
\end{equation}
The same exponent holds for local-stochastic record noise: if
\begin{equation}
 \Pr[S\subseteq\operatorname{supp}(e)]\le(Kq)^{|S|}
\end{equation}
for every support set $S$, a union bound over the finite set of nonzero valid-code displacements gives $P_{\rm wv}=O(q^d)$.

For linear schedules, the binary Griesmer bound gives
\begin{equation}
 n\ge\sum_{j=0}^{k-1}\left\lceil\frac{d}{2^j}\right\rceil.
\end{equation}
The repeated-simplex family used in the Letter has parameters
\begin{equation}
 [n,k,d]=[r(2^k-1),k,r2^{k-1}],
\end{equation}
and reaches both the Griesmer bound and the unrestricted binary Plotkin bound when $d=r2^{k-1}$. Thus the Clifford-parity restriction requires no additional measurements in the distance-four $T$, $CS$, and $CCZ$ examples.

\section{Optimal schedules for two branch bits}
For $k=2$, there are three nonzero check labels: $10$, $01$, and $11$. Let their multiplicities be $n_{10}$, $n_{01}$, and $n_{11}$. The three nonzero codeword weights are
\begin{align}
 w_{10}&=n_{10}+n_{11},\\
 w_{01}&=n_{01}+n_{11},\\
 w_{11}&=n_{10}+n_{01}.
\end{align}
Writing $n=n_{10}+n_{01}+n_{11}$,
\begin{equation}
 d=n-\max\{n_{10},n_{01},n_{11}\}.
\end{equation}
Therefore $d\le\lfloor2n/3\rfloor$, or
\begin{equation}
 n\ge\left\lceil\frac{3d}{2}\right\rceil.
\end{equation}
The global bound in the preceding section gives the same inequality without assuming linearity. For $d=2r$, multiplicities $(r,r,r)$ are optimal. For $d=2r+1$, $(r+1,r+1,r)$ and permutations are optimal. Thus $[5,2,3]$ and $[6,2,4]$ are globally shortest binary records for four branches at distances three and four, respectively.

\section{$CS$ checks and measurement-record error probabilities}
For $U=CS_{AB}$,
\begin{align}
 C_1&=X_A S_B CZ_{AB},\\
 C_2&=X_B S_A CZ_{AB},\\
 C_{12}&=C_1C_2=H_{XY}^{(A)}H_{XY}^{(B)},
\end{align}
where $H_{XY}=(X+Y)/\sqrt2$. The five-check and six-check schedules are
\begin{align}
 &(C_1,C_2,C_{12},C_1,C_2),\\
 &(C_1,C_2,C_{12},C_1,C_2,C_{12}),
\end{align}
with weight enumerators
\begin{align}
 W_5(z)&=1+2z^3+z^4,\\
 W_6(z)&=1+3z^4.
\end{align}
Thus, for independent logical record-bit flips of probability $q$,
\begin{align}
 P_{\rm wv}^{(5)}&=2q^3(1-q)^2+q^4(1-q),\\
 P_{\rm wv}^{(6)}&=3q^4(1-q)^2.
\end{align}
For comparison, independently repeating $C_1$ and $C_2$ four times gives a $[8,2,4]$ record with weight enumerator
\begin{equation}
 W_8(z)=1+2z^4+z^8,
\end{equation}
and hence
\begin{equation}
 P_{\rm wv}^{(8)}=2q^4(1-q)^4+q^8.
\end{equation}
Both six- and eight-check records are quartic. The coded record trades a modestly larger leading wrong-valid coefficient ($3$ rather than $2$) for two fewer logical measurements. Because wrong-valid events are already fourth order, the shorter record can simultaneously increase the valid-record probability. In the CAT7 readout model $q=[1-(1-2p_m)^7]/2$ at $p_m=10^{-3}$, $P_{\rm valid}^{(6)}=0.9589707$ and $P_{\rm valid}^{(8)}=0.9456718$, while $P_{\rm wv}^{(6)}=6.93\times10^{-9}$ and $P_{\rm wv}^{(8)}=4.56\times10^{-9}$. These are record-only quantities, not full-factory logical error rates.

\section{Physical cost of the $CS$ distance-four schedules}
The logical measurement-count saving survives a concrete native-CZZ-assisted Steane compilation. The physical gate counts for the five- and six-check schedules are, respectively,
\begin{align}
 L_5&=(35\ {\rm CZZ},\ 56\ {\rm CX},\ 196\ {\rm 1Q}),\\
 L_6&=(42\ {\rm CZZ},\ 56\ {\rm CX},\ 224\ {\rm 1Q}).
\end{align}
The five-check schedule contains two $C_1$, two $C_2$, and one $C_{12}$ measurements, whereas the six-check schedule contains two of each. Their difference therefore gives the physical gate inventory of one $C_{12}$ check,
\begin{equation}
 L(C_{12})=(7,0,28),\qquad |L(C_{12})|=35.
\end{equation}
By the $A\leftrightarrow B$ symmetry, half of $L_6/2-L(C_{12})$ is the inventory of either $C_1$ or $C_2$,
\begin{equation}
 L(C_1)=L(C_2)=(7,14,42),\qquad |L(C_1)|=|L(C_2)|=63.
\end{equation}

Each logical check uses one verified CAT7 ancilla. Its preparation and verification contain 11 preparations, 13 CX gates, and four verifier measurements, followed by seven CAT readout measurements, for 35 additional active locations per check. Thus the total costs per logical check before the midpoint detector are
\begin{center}
\begin{tabular}{c c c c}
\toprule
check & check gates & CAT7 & total \\
\midrule
$C_1$ & 63 & 35 & 98 \\
$C_2$ & 63 & 35 & 98 \\
$C_{12}$ & 35 & 35 & 70 \\
\bottomrule
\end{tabular}
\end{center}
The full midpoint Steane error-detection step costs 82 active locations per logical block, hence 164 for two blocks.

An independent distance-four record repeats $C_1$ and $C_2$ four times each. Its measurement stage therefore costs
\begin{equation}
 L_{\rm rep}=4(98)+4(98)=784,
\end{equation}
and including the fixed midpoint detector gives
\begin{equation}
 L_{\rm rep}^{\rm core}=784+164=948.
\end{equation}
The optimal $[6,2,4]$ schedule uses two measurements of each parity,
\begin{equation}
 L_{\rm code}=2(98+98+70)=532,
\end{equation}
so
\begin{equation}
 L_{\rm code}^{\rm core}=532+164=696.
\end{equation}
Consequently coding reduces the measurement-stage count by $32.1\%$ and the cultivation-core operation count by
\begin{equation}
 1-\frac{696}{948}=26.6\%.
\end{equation}
These ``active locations'' are noisy circuit-operation counts. They are not spacetime volume, do not include encoded-state preparation or the subsequent code escape, and should not be interpreted as a complete factory-resource reduction. Their role here is narrower: they verify that record-code compression remains a substantial physical saving after unequal check costs and a fixed error-detection overhead are included.

\subsection{Qubit-step cost cross-check}
As an independent resource cross-check, we also count active-qubit gate steps under the same local scheduling convention; idle/storage intervals are not charged. Each verified CAT7 preparation/verification costs $150$ qubit-steps. The check-body-plus-readout costs are $168$, $168$, and $84$ qubit-steps for $C_1$, $C_2$, and $C_{12}$, respectively, and the common two-block midpoint detector costs $300$. Hence the coded core requires
\begin{equation}
 Q_6=6(150)+2(168+168+84)+300=2040,
\end{equation}
whereas independent quartic repetition requires
\begin{equation}
 Q_8=8(150)+8(168)+300=2844.
\end{equation}
Thus the same schedule that saves $26.6\%$ of active locations saves
\begin{equation}
 1-Q_6/Q_8=28.27\%
\end{equation}
of this independent qubit-step proxy. Including the common encoded-preparation and initial-detection front end ($356$ qubit-steps) gives $2396$ versus $3200$, a $25.13\%$ reduction. These quantities are not hardware spacetime costs: they deliberately omit idle/storage cost and are used only as a cross-metric check that the active-location advantage is not an artifact of one resource-counting convention.

The value $532$ is in fact the global minimum over all distance-four Clifford-parity multiplicities in this compiled cost model. Writing $(m_1,m_2,m_{12})$ for the numbers of $C_1,C_2,C_{12}$ checks, respectively, the three nonzero branch differences impose
\begin{equation}
 m_1+m_{12}\ge4,\qquad m_2+m_{12}\ge4,\qquad m_1+m_2\ge4.
 \label{eq:weightedCSconstraintsS}
\end{equation}
The cost function is
\begin{equation}
 \mathcal L=98(m_1+m_2)+70m_{12}.
\end{equation}
Direct enumeration gives the unique optimum $(m_1,m_2,m_{12})=(2,2,2)$ and $\mathcal L=532$. A closed-form proof for general distance is given below.

\section{$CCZ$ logical checks}
For $U=CCZ_{123}$,
\begin{align}
 C_1&=CCZ\,X_1\,CCZ=X_1CZ_{23},\\
 C_2&=X_2CZ_{13},\\
 C_3&=X_3CZ_{12}.
\end{align}
The seven columns
\begin{equation}
 100,010,001,110,101,011,111
\end{equation}
form a $[7,3,4]$ simplex generator matrix. Every nonzero message has weight four. The global binary bound for $k=3,d=4$ gives
\begin{equation}
 n\ge\left\lceil4\frac{7}{4}\right\rceil=7,
\end{equation}
so this schedule is shortest even if nonlinear record assignments were allowed.

\section{Exact $CCZ$ measurement circuit and implementation scope}
The record-code theorem specifies \emph{which} logical involutions should be measured; it does not claim that every such measurement has the same physical cost. It is nevertheless useful to separate exact measurement semantics from fault-tolerant compilation.

For any Hermitian operator $C=C^\dagger$ satisfying $C^2=I$, prepare an ancilla in $|+\rangle$, apply a controlled-$C$, and measure the ancilla in the $X$ basis. If the outcome bit is $m\in\{0,1\}$, the unnormalized data state is
\begin{equation}
 \Pi_m|\psi\rangle,\qquad
 \Pi_m=\frac{I+(-1)^m C}{2}.
 \label{eq:projectorSemantics}
\end{equation}
Thus this circuit realizes the exact nondemolition projective measurement associated with the record bit.

\begin{figure}[h]
\centering
\includegraphics[width=0.6\linewidth]{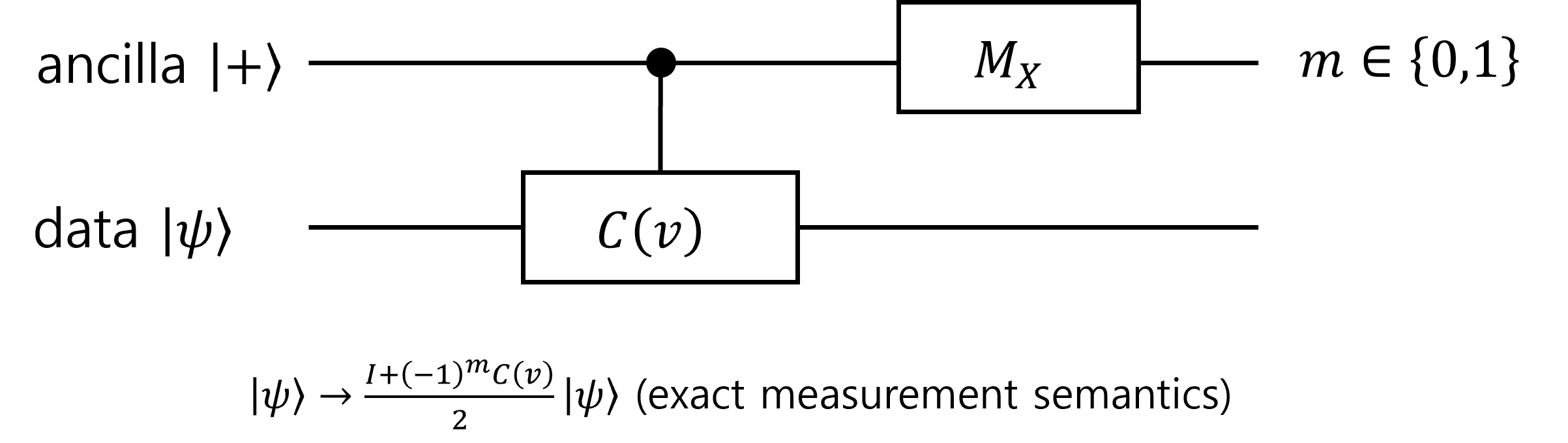}
\caption{Exact semantics of one logical record measurement. The circuit realizes the projectors in Eq.~(\ref{eq:projectorSemantics}); it is an ideal reference circuit, not a claim of low fault-tolerant overhead.}
\label{fig:semantic}
\end{figure}

For the first $CCZ$ symmetry,
\begin{equation}
 C_1=X_1CZ_{23},
\end{equation}
the controlled operation in Fig.~\ref{fig:semantic} factorizes exactly as
\begin{equation}
 \operatorname{ctrl}(C_1)
 =\operatorname{CNOT}_{a\rightarrow1}\,CCZ_{a23}.
 \label{eq:controlledCCZcheck}
\end{equation}
Equation~(\ref{eq:controlledCCZcheck}) is intentionally \emph{not} proposed as a fault-tolerant implementation using only stabilizer operations: it exposes that a generic exact controlled-measurement construction can itself require a third-level primitive. Clifford closure of $C(v)$ therefore does not imply a stabilizer-only measurement gadget.

Existing gauging constructions provide low-overhead fault-tolerant logical measurement for Pauli operators in arbitrary stabilizer codes~\cite{WilliamsonYoder2026}, and recent cultivation work extends gauging to particular non-Pauli transversal Hermitian Clifford symmetries of the color code~\cite{Hetenyi2026}. These results establish that non-Pauli logical Clifford measurement can be fault tolerant in suitable encodings, but they do not automatically furnish a compiled gadget for the multiblock $X_jCZ_{\bar j}$ family above. The $[7,3,4]$ result should therefore be read as a globally minimal \emph{logical measurement schedule}; circuit distance, ancilla connectivity, and additional non-Clifford gates in the physical implementation remain architecture-specific compilation questions.

\section{Minimum-cost logical-check schedules}
Equal check count is not always the correct physical objective because distinct logical Clifford observables may have different gadget depths, ancilla requirements, or entangling-gate counts. Let $m_v$ be the number of times parity $v\ne0$ is measured and let $c_v>0$ denote its cost. Branch difference $a\ne0$ changes precisely the checks with $a\cdot v=1$, so distance at least $d$ is equivalent to
\begin{equation}
 \sum_{v:a\cdot v=1}m_v\ge d
 \quad\text{for every }a\in\F_2^k\setminus\{0\}.
\end{equation}
The minimum-cost schedule is therefore the following integer optimization problem
\begin{align}
 \min_{m_v\in\mathbb Z_{\ge0}}\quad &\sum_{v\ne0}c_vm_v,\\
 \text{subject to}\quad&\sum_{v:a\cdot v=1}m_v\ge d
 \quad \forall a\ne0.
\end{align}
For $c_v=1$ this reduces to minimum-length linear record design.

For $k=2$, suppose $c_{10}=c_{01}=A$ and $c_{11}=B$, and write $(x,y,z)=(m_{10},m_{01},m_{11})$. The constraints are
\begin{equation}
 x+z\ge d,\qquad y+z\ge d,\qquad x+y\ge d.
\end{equation}
Set $s=x+y$. At fixed $s$, the smallest admissible $z$ is obtained by balancing $x$ and $y$, which maximizes $\min(x,y)=\lfloor s/2\rfloor$; hence
\begin{equation}
 z_{\min}(s)=d-\lfloor s/2\rfloor,
 \qquad d\le s\le2d.
\end{equation}
The minimum cost at fixed $s$ is therefore
\begin{equation}
 F(s)=As+B\bigl[d-\lfloor s/2\rfloor\bigr].
 \label{eq:weightedFs}
\end{equation}
For the even target distances $d=2r$ relevant here, an odd value of $s$ is never optimal because increasing an even $s$ by one leaves $\lfloor s/2\rfloor$ unchanged while adding cost $A$. Restricting to even $s$, increasing $s$ by two changes $F$ by $2A-B$. Consequently,
\begin{equation}
 C_{\min}(2r)=
 \begin{cases}
 r(2A+B),& B<2A,\\
 4Ar,& B>2A,
 \end{cases}
 \label{eq:weightedclosedS}
\end{equation}
with a family of degenerate even-$s$ optima at $B=2A$. For $B<2A$ the unique optimum is $(x,y,z)=(r,r,r)$. In the compiled $CS$ model $A=98$, $B=70<196$, and $d=4$ ($r=2$), so Eq.~(\ref{eq:weightedclosedS}) gives uniquely
\begin{equation}
 (x,y,z)=(2,2,2),\qquad C_{\min}=532.
\end{equation}
An independent exhaustive integer search for all even $d=2,4,\ldots,20$ was used as a deterministic implementation check of Eq.~(\ref{eq:weightedclosedS}).

\section{Circuit-level comparison at the Steane output}
To test whether the shorter weighted-optimal schedule incurs a physical acceptance or output-error penalty, we extended the preserved 14-data-qubit exact state-vector simulator to the independent eight-check record. The two schedules are
\begin{align}
 \text{coded: }& C_1,C_2,C_{12}\mid D_{\rm mid}\mid C_1,C_2,C_{12},\\
 \text{repeat: }& C_1,C_2,C_1,C_2\mid D_{\rm mid}\mid C_1,C_2,C_1,C_2.
\end{align}
For both schedules the simulator applies the explicit physical compilation of each check on seven Steane coordinates. A one-qubit location is followed, upon failure, by a uniformly selected nonidentity Pauli; a CX by one of 15 nonidentity two-qubit Paulis; and a CZZ by one of 63 nonidentity ancilla--data--data Pauli patterns. The ancilla component flips the logical check readout according to the preserved model, and each logical record bit also receives the parity of seven independent physical readout flips.

The midpoint Steane boundary is an ideal code-space projection, while the final ideal code-space projection is removed. After an accepted record, the state is instead classified by sequential projective measurement of the six Steane stabilizers on each logical block, yielding a two-block Steane syndrome-sector pair. For each sector we evaluate the weight of the frozen canonical Pauli representative; this weight is an output-error diagnostic only, not a decoder-distance claim. CAT7 preparation/verification faults, a noisy physical midpoint-ED circuit, encoded-state preparation, and all downstream surface-code escape faults remain outside this comparison. These omissions are common to the two schedules and prevent the data below from being interpreted as full-factory logical-error estimates.

Five thousand shots were used for every protocol and $p$:
\begin{center}
\begin{tabular}{c cc cc}
\toprule
& \multicolumn{2}{c}{acceptance without final projection} & \multicolumn{2}{c}{mean residual Pauli weight}\\
$p$ & coded & repeat & coded & repeat\\
\midrule
$2\times10^{-4}$ & $0.9366(34)$ & $0.9084(41)$ & $0.0130(20)$ & $0.0295(33)$\\
$5\times10^{-4}$ & $0.8574(49)$ & $0.7844(58)$ & $0.0425(40)$ & $0.0752(58)$\\
$10^{-3}$ & $0.7350(62)$ & $0.6256(68)$ & $0.0639(54)$ & $0.1263(84)$\\
\bottomrule
\end{tabular}
\end{center}
Parentheses denote one-standard-error uncertainties. At $p=10^{-3}$, the fraction of accepted shots for which both Steane blocks are already in the zero-syndrome sector is $0.95837$ for the coded schedule and $0.92327$ for repetition. Thus removing the final ideal projection does not reveal a hidden syndrome penalty at the Steane output: over the tested range the six-check schedule has both higher acceptance without final projection and a smaller mean residual Pauli weight.

\begin{figure}[h]
\centering
\includegraphics[width=0.47\linewidth]{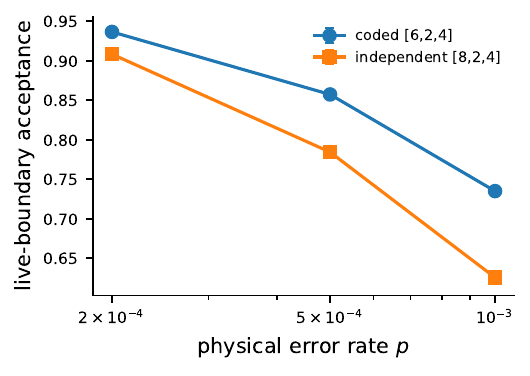}\hfill
\includegraphics[width=0.47\linewidth]{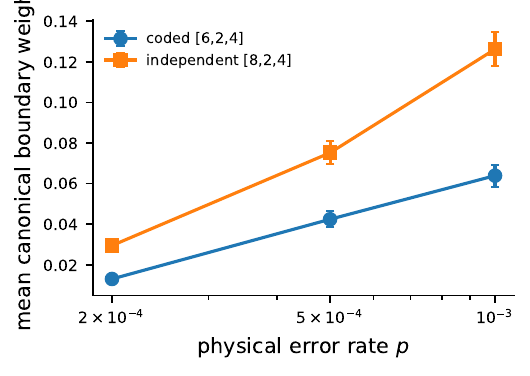}
\caption{Live Steane-boundary diagnostics after removing the final ideal code-space projection. Left: record-and-midpoint acceptance. Right: mean weight of the canonical two-block Steane boundary representative among accepted shots. Error bars show one standard error. These are boundary-quality diagnostics, not full-factory logical-error estimates.}
\label{fig:liveBoundary}
\end{figure}

A current full-factory comparison would additionally require regenerating the downstream boundary distributions and decoder priors for the eight-check schedule. We therefore do not transplant the six-check downstream decoder to the repetition baseline or claim a downstream logical-error comparison from these data. The Steane-output test isolates the statement needed here: record compression improves acceptance without delivering a dirtier Steane boundary to the subsequent stage.

\section{Record-error certificate and a third-order data-fault example}
For each elementary readout fault whose sole logical effect is to flip one record coordinate, define its fault signature $s(f)=e_t$, where $t$ is the affected logical check. A set $F$ of such faults has signature
\begin{equation}
 s(F)=\bigoplus_{f\in F}s(f).
\end{equation}
For the $[6,2,4]$ record, every nonzero valid-code displacement has weight four. Therefore no set of at most three single-coordinate record-only faults can produce a wrong valid record. Exhaustive enumeration of the $\binom61+\binom62+\binom63=41$ nonzero signatures of weight at most three gives zero signatures that map to a different valid branch. This certificate covers elementary parity-readout faults localized to one logical check; it does not cover faults that simultaneously alter quantum data and reported outcomes.

This distinction is physically necessary. In the same compiled late check window $(C_1,C_2,C_{12})$, the exact state-vector model reproduces an explicit accepted third-order data-fault mechanism on the $(a,b)=(0,0)$ branch:
\begin{enumerate}
\item $C_1$, Steane coordinate 0, CZZ location: $I_{\rm anc}Y_A I_B$;
\item $C_2$, coordinate 1, first wrapper CX: $Y_A Z_B$;
\item $C_{12}$, coordinate 2, post-CZZ $A$-side one-qubit location: $Z_A$.
\end{enumerate}
Holding the valid observed late record at $000$, the final code-space projection has probability $1-1.3\times10^{-14}$ and the normalized state has fidelity $1-4\times10^{-16}$ with $Z_A|CS\rangle$ and essentially zero fidelity with $|CS\rangle$. Thus the full physical channel can remain cubic even though the record-only channel is quartic. The role of the coded record is to remove record corruption from the leading cubic bottleneck, not to promote the complete factory to fourth order.

\section{Separation of measurement-record and data faults}
The record distance controls only errors that corrupt the logical classical outcomes while leaving the intended branch relation otherwise well defined. Faults inside the gadget that simultaneously change the data and its reported outcome can evade a pure record-code argument. Consequently, the main-text statement $P_{\rm wv}=O(q^d)$ must not be promoted to an unconditional full-factory distance theorem.

A useful composability rule is nevertheless immediate. Suppose the accepted data/decoder channel begins at order $p^\delta$, while record noise satisfies $q=\Theta(p)$. Then the two mechanisms enter at orders $p^\delta$ and $p^d$, respectively, so the leading exponent is at most $\min(\delta,d)$ and the measurement-record contribution is higher order in $p$ whenever $d\ge\delta+1$. If $d=\delta$, the record layer contributes to the same leading coefficient and may dominate it if its multiplicity is large. The explicit cubic data witness above together with the quartic record certificate illustrates how the leading error mechanism shifts from readout corruption to data faults for the direct-$CS$ core.